# PARITY OF THE SOLAR MAGNETIC FIELDS AND RELATED ASTROPHYSICAL PHENOMENA

Gopkumar,G    and    Girish,T.E
Department of Physics,University College,
Trivandrum  695 034, Kerala
Email:  tegirish5@yahoo,com

## Abstract

The cumulative contribution of odd ($B_o$) and even ($B_E$) parity zonal magnetic multipoles to the solar magnetic fields is calculated using spherical harmonic coefficients of the photospheric magnetic field for the years 1959-1985. The dominant parity of the solar magnetic field is shown to change from odd to even during every sunspot cycle. The association of variations of $B_o$ and $B_E$ with different astrophysical phenomena such as magnetic reversal of solar polar magnetic fields, north-south asymmetry in sunspot activity and strength of the interplanetary magnetic field will be also discussed. Using solar observations we could infer that dominant parity of the solar magnetic field is changing from even to odd during the past 12 solar cycles when the solar activity is showing an increasing trend during this period.

## 1. Introduction

The study of sun and its influence on the heliosphere is helpful in understanding the environment of other stars with planetary systems sustaining life [1]. The physical environment near earth and the solar system is basically controlled by the time variations of the solar output in the form of electromagnetic radiation, particles ( solar wind) and weak interplanetary magnetic fields ( IMF). Solar output changes is closely related to variations of solar magnetic fields which is manifested in the different forms like the the well known sunspot activity whose telescopic observations started four hundred years ago. Direct and systematic observations of the solar magnetic fields started only during late 1950's coinciding with the beginning of the space age. However past solar magnetic phenomena can be inferred from other solar and planetary observations [2].

Parity is an important concept in modern physics.The role of parity in understanding time variations of solar magnetic fields is first brought to focus by Prof.Stenflo. through spherical harmonic analysis of solar magnetic fields using Mt.Wilson and Kitt Peak solar observations [3,4]. Gokhale and Javariah [5] extended these studies back to the 19$^{th}$ century using sunspot data. In this paper we have studied cyclic and long term changes in the dominant parity of the solar magnetic fields during the sunspot cycles 11-23 ( 1867-2007) in relation with different solar-terrestrial phenemena such as magnetic reversal of solar polar magnetic fields, north-south asymmetry in the sunspot activity and variations in the IMF strength observed near 1 AU. This work is in continuation of earlier efforts in the similar direction [6-8].

## 2. Variations of the cumulative contributions of zonal magnetic multipoles of odd and even parity to the solar magnetic fields during the years 1959-1985 and related solar terrestrial phenomena

(a) The cumulative (net) contributions of the axisymmetric or zonal solar magnetic multipoles of odd parity ( magnetic dipole symmetry) to the photospheric magnetic field can be defined by the relation

$$Bo = SQRT ( \sum_{l=odd} g_l^2 ) \qquad (1)$$

Here l= 1,3,5,7, 9,11 and 13 and $g_l$ is the axisymmetric apherical harmonic multipole of order l.

(b) The cumulative ( net) contributions of the axisymmetric or zonal magnetic multipoles of even parity ( magnetic quadrupole symmetry) to the photospheric magnetic field can be defined by the relation

$$B_E = \text{SQRT}\left(\sum_{l=\text{even}} g_l^2\right) \qquad (2)$$

Here l=2,4,6,8,10,12 and 14.

(c) The ratio of the even parity contributions to the odd parity contributions of the zonal solar magnetic field is given by the relation

$$\alpha = B_E/Bo \qquad (3)$$

Using zonal spherical harmonic coefficients up to the order 14 [9] of the solar magnetic field we have calculated the values of Bo, $B_E$ and α for the Carrington rotation periods 1417-1761 covering years 1959-1985. The results are shown in Fig 1.

The zonal even parity contributions $B_E$ show a systematic change during a sunspot cycle with a minimum value during sunspot minima and maximum value during the polar reversal periods. During the solar cycle 21, $B_E$ is found reach a maximum value during the CR 1707 coinciding with the final magnetic polarity reversal of the solar polar magnetic fields.
Further the magnetic monopole situation of the solar polar magnetic field ( north and south poles with same magnetic polarity) is a reflection of the dominance of the zonal even parity contribution over the zonal odd parity contribution. During such occasions α > 1 which is observed during CR 1705-1707 before the final polarity reversal of the solar magnetic fields. The variations of the yearly mean photospheric magnetic flux ( Fig 2) during the years 1966-1996 is found to show correlated changes with the $B_E$ variations during the same period. The variations of the yearly north-south asymmetry ($\Delta_s$) of the sunspot activity during the years ( Fig 3) is however observed to be similar to that of α.

The weak interplanetary magnetic field is basically the large scale solar magnetic field carried by the solar wind. The yearly means of the total field strength ( B) or intensity of the IMF observed/inferred near earth during the years 1959-1985 is plotted in Fig 4. It is interesting observe that IMF B show a better correlation with Bo variations during the same period.

### 3. Long term changes in the dominant parity of the solar magnetic field and its consequences

In Table 1 we have given the epochs of reversal of solar polar magnetic fields during sunspot cycles 11-23 inferred from $H_\alpha$ synoptic or solar magnetograph observations [10] Due to the differences in the timings of reversal of solar polar fields in the northern and southern heliohemispheres magnetic monopole type situation generally prevails during these epochs. As found in the previous section this will correspond to situation where we have $\alpha > 1$ or the dominant polarity of the solar magnetic field is even with magnetic quadrupole symmetry. The duration of the solar polar reversal for each cycle is also shown in this Table which is found to be smaller for recent sunspot cycles compared to the cycles 12 and 14.

In Fig 5 we have plotted sunspot cycle averaged values of the absolute ( sign independent) north-south asymmetry parameter of the sunspot activit , $|\Delta_s|$ . This is calculated using yearly north-south asymmetry in sunspot area during the years 1874-1946 and in sunspot number for the years 1947-1984. The values of mean IMF B which is inferred or observed near 1 AU during these sunspot cycles is also plotted in this figure for comparison. We can find a clear decreasing trend in $|\Delta_s|$ between the sunspot cycles 12-21 when we can observe an increasing trend for IMF B during this period. The long term decreasing trend in magnitude of north-south asymmetry in sunpot activity can be explained in terms of similar decrease in $\alpha$ or the relative dominance of even parity over odd parity in the zonal solar magnetic field.

### 4. Discussion

We have found that the dominant polarity of the solar magnetic field exhibit long term changes apart from variations within a sunspot cycle. The dominant parity is odd ( magnetic dipole symmetry) during the sunspot minima which changes to even ( magnetic quadrupole symmetry) during the sunpot maxima in general and solar polar reversal period in particular. This result is better illustrated in Fig 5 where even parity contributions ($B_E$) is plotted along with the north-south dipole ( odd parity component of the first order) harmonic $(g_1)^2$. The even to odd parity contributions ratio ( $\alpha$ ) reaches a maxima during the solar polar magnetic reversal epochs. Odd parity contributions are found to be correlated with interplanetary magnetic field strength variations near 1 AU and $\alpha$ changes are correlated with variations of the magnitude of north-south asymmetry in sunspot activity observed during the years 1959-1985.

Based upon the results obtained during the recent sunspot cycles we have attempted to infer long term changes in the dominant parity of the solar magnetic field using solar observations during the sunspot cycles 11-23 ( 1867-2007). The cyclic averages of the magnitude of north-south asymmetry in sunspot activity and IMF B during these cycles is found to show anticorrelated variations. We interpret this result that the odd parity contributions to the solar magnetic field is increasing during the past 12 solar cycles along with solar activity while the even parity contributions is probably decreasing. This is further supported by the decrease in the duration of solar polar magnetic reversal period on an average from the late 19$^{th}$ century to the beginning of the 21$^{st}$ century. The relation of this long term parity change of solar magnetic field with the observed decrease in solar magnetic flux amplification factor during solar cycles 9-23 [11] will be addressed in a future publication.

**Aknowledgements**

One of the authors ( G.Gopkumar) wish to thank UGC for the award of a teacher research fellowship under the FIP programme to compleate the Ph.D thesis work in University of Kerala when part of the present paper is completed. The authors are also grateful to Prof.K.R,Sivaraman of IIA,Bangalore for his valuable efforts to infer solar polar reversal epochs back to the 19$^{th}$ century .

**Table 1. Epochs of magnetic polarity reversals in the polar regions of the sun during sunspot cycles 11-23 and dominant parity of the solar magnetic field**

| Sunspot cycle | Epochs of polar magnetic reversal | Duration in years ($\alpha > 1$) |
|---|---|---|
| 11 | 1872.3 | 0 |
| 12 | 1883.4 to 1885.8 | 2.4 |
| 13 | 1895.0 | 0 |
| 14 | 1905.3 to 1908.4 | 3.1 |
| 15 | 1918.6 to 1918.7 | 0.1 |
| 16 | 1927.9 to 1929.9 | 2 |
| 17 | 1940.0 to 1940.1 | 0.1 |
| 18 | 1949.0 to 1950.2 | 1.1 |
| 19 | 1958.0 to 1959.7 | 1.7 |
| 20 | 1969.7 to 1971.1 | 1.6 |
| 21 | 1981.0 to 1981.8 | 0.8 |
| 22 | 1990.8 to 1991.8 | 1.0 |
| 23 | 2002.0 to 2002.8 | 0.8 |

**Figure Captions**

Fig 1. Smoothed values of the cumulative contribution of zonal odd parity magnetic multipoles ($B_o$) to the photospheric magnetic field during Carrington rotations 1417-1761, during the years 1959-1985.

Fig 2. Smoothed values of the cumulative contribution of zonal even parity magnetic multipoles ($B_E$) to the the photospheric magnetic field during the Carrington rotations 1417-1761, during the years 1959-1985.

Fig 3. Smoothed values of the ratio $B_E/B_o$ ( $\alpha$ ) during the Carrington rotations 1417-1761 ,during the years 1959-1985.

Fig 4. Smoothed yearly means of the total photospheric magnetic flux observed during the years 1966-96.

Fig.5. Magnitude of the yearly north-south asymmetry in sunspot activity during the years 1959-84.

Fig.6. Yearly mean values of the IMF strength inferred/observed near 1 AU during the years 1959-85.

Fig.7 Sunspot cycle averaged magnitude of the north-south asymmetry in sunspot activity and strength of the interplanetary magnetic field near 1 AU during the sunspot cycles 11-22.

Fig 8 Yearly values of the net zonal even partity contributions to the solar magnetic field ($B_E$) and north-south dipolar contribution to the solar magnetic field ( $g_1$ )$^2$ during the Carrington rotations 1417-1761, during the years 1959-1985.

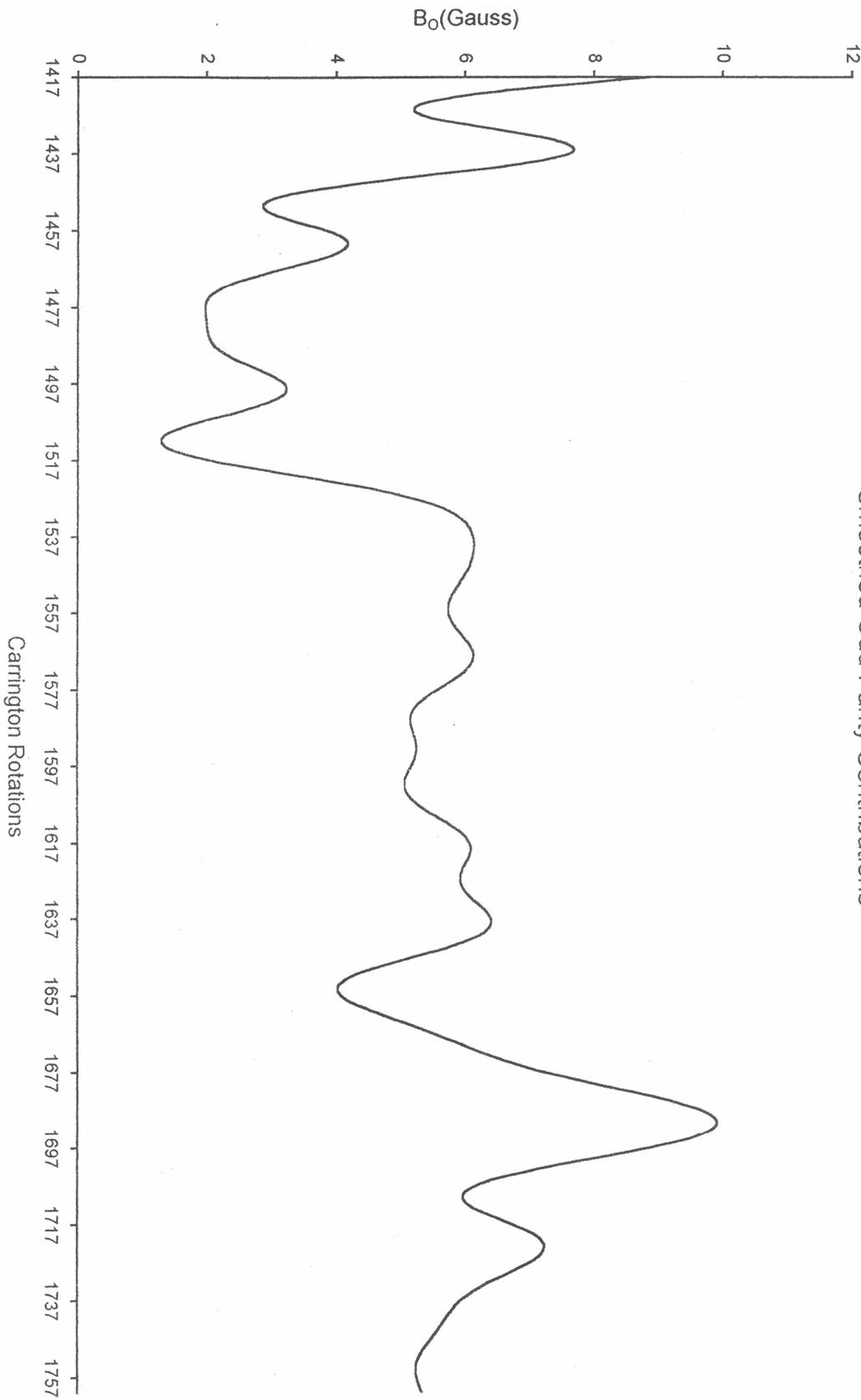

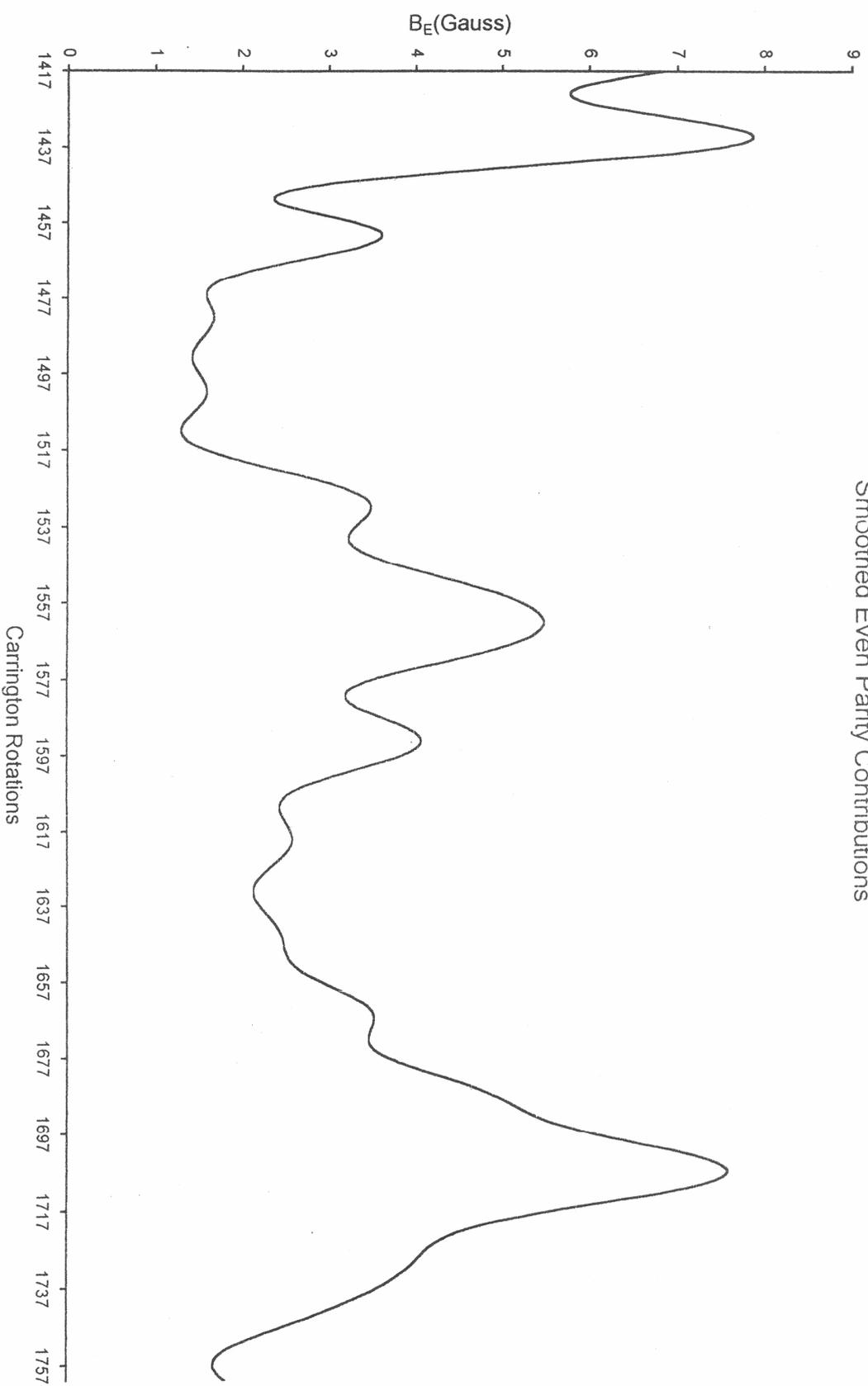

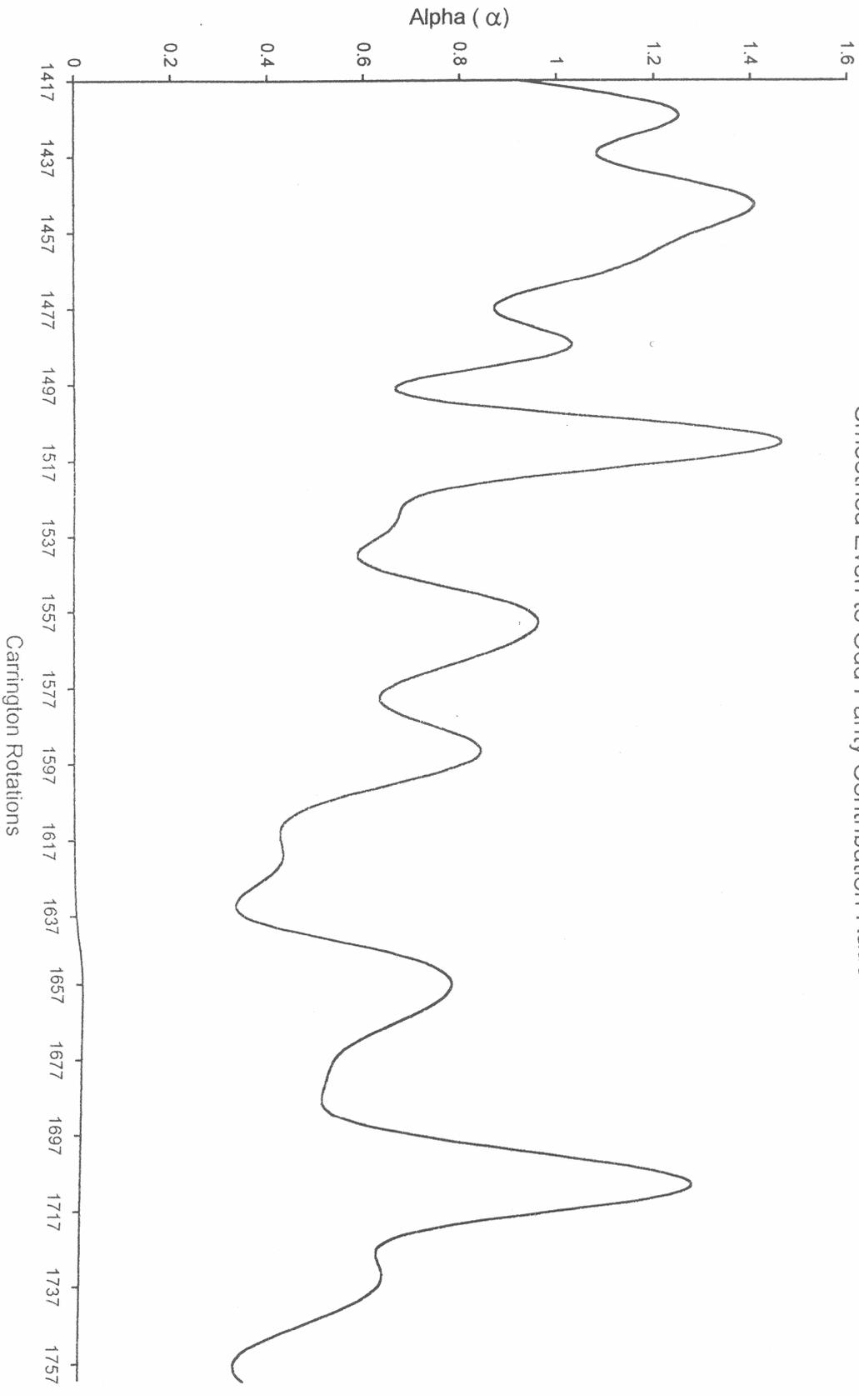

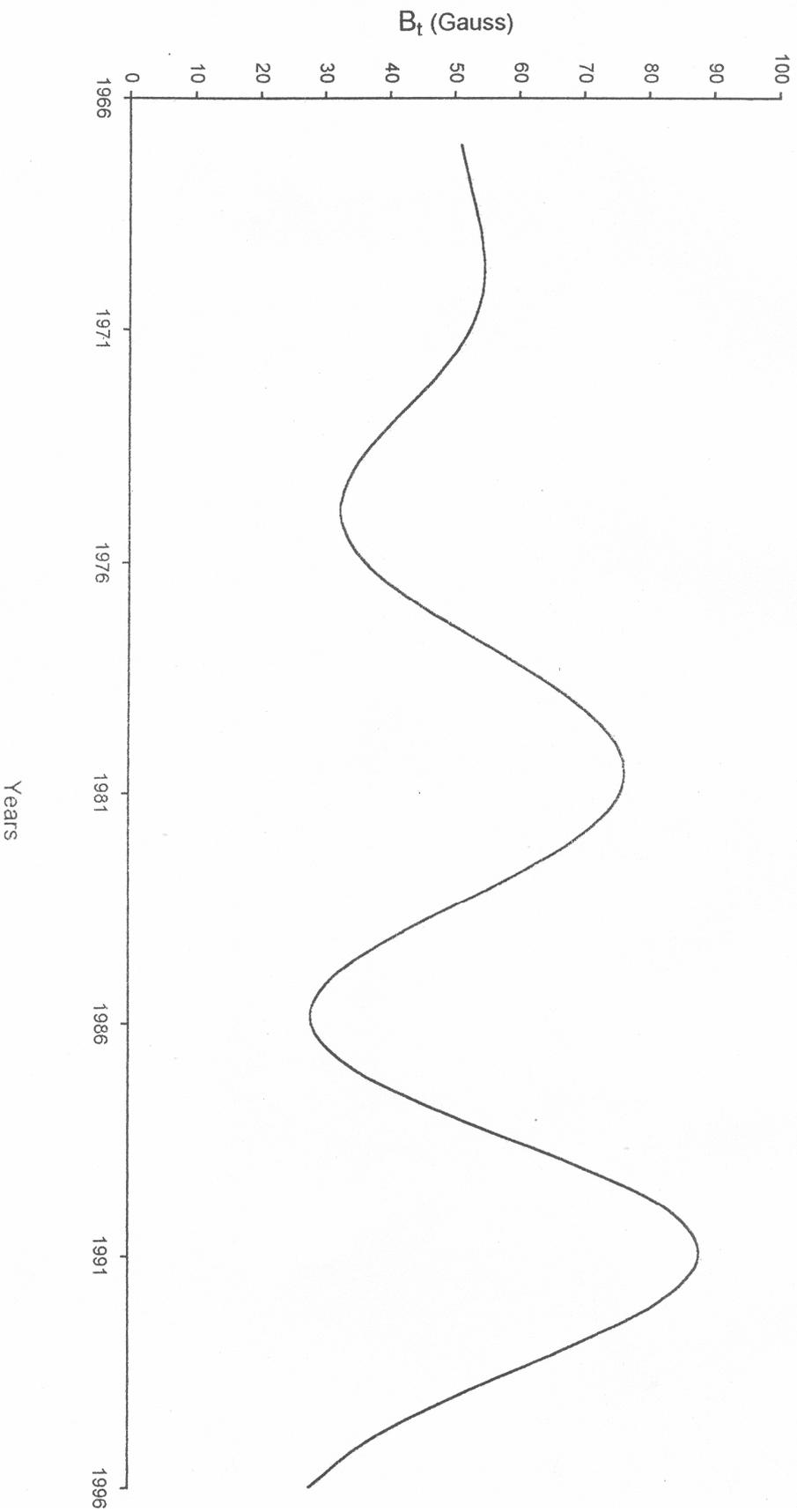

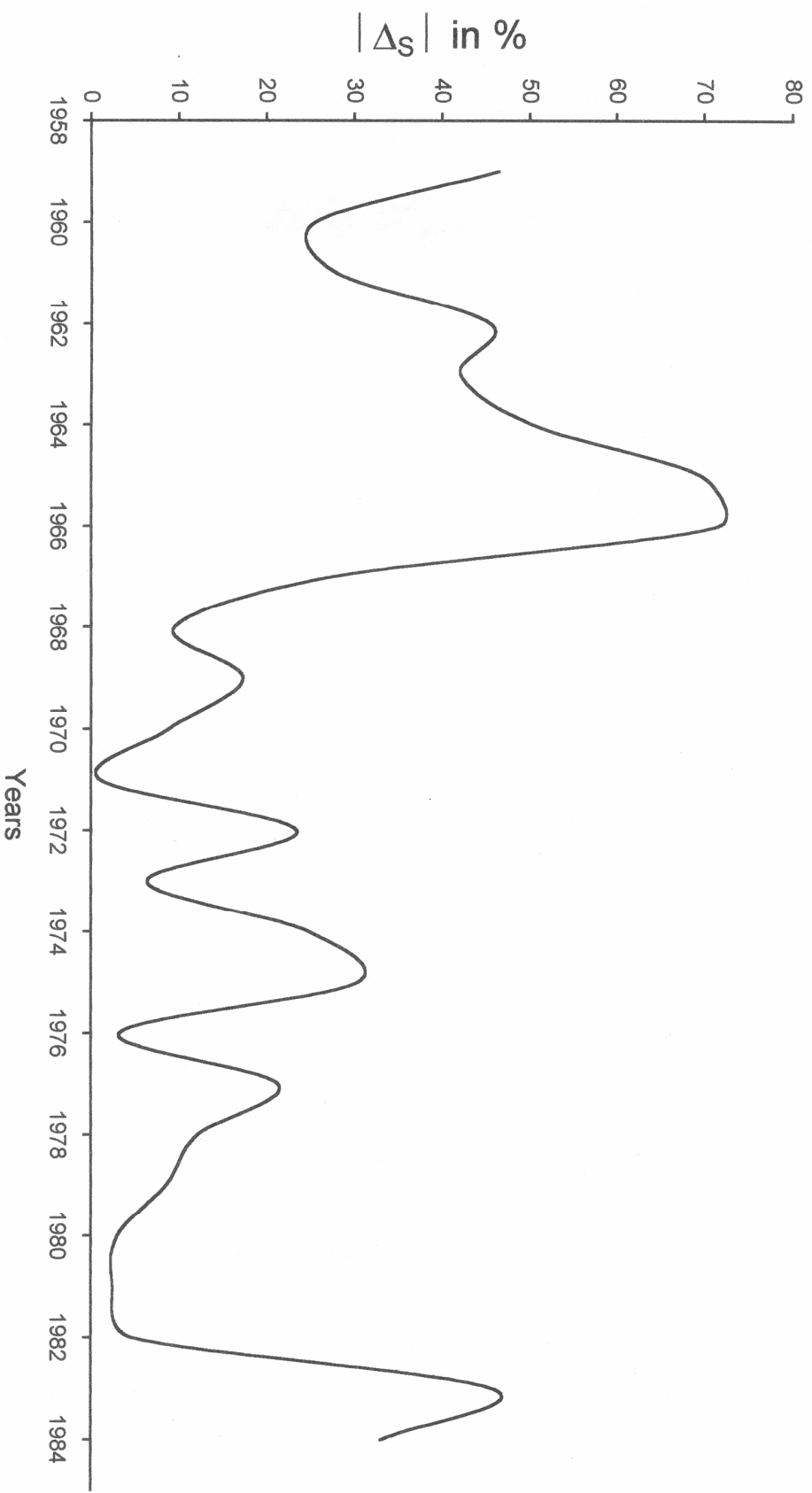

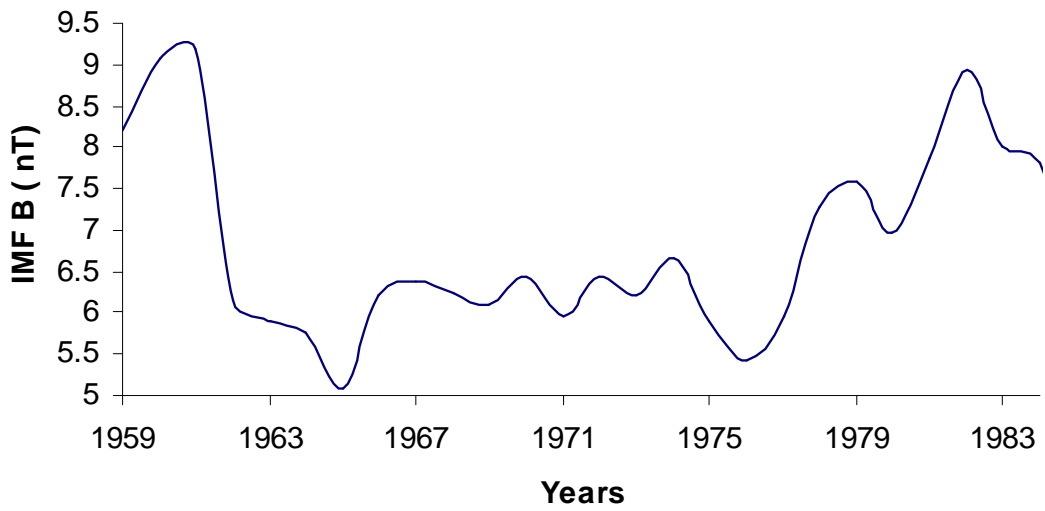

Fig 6

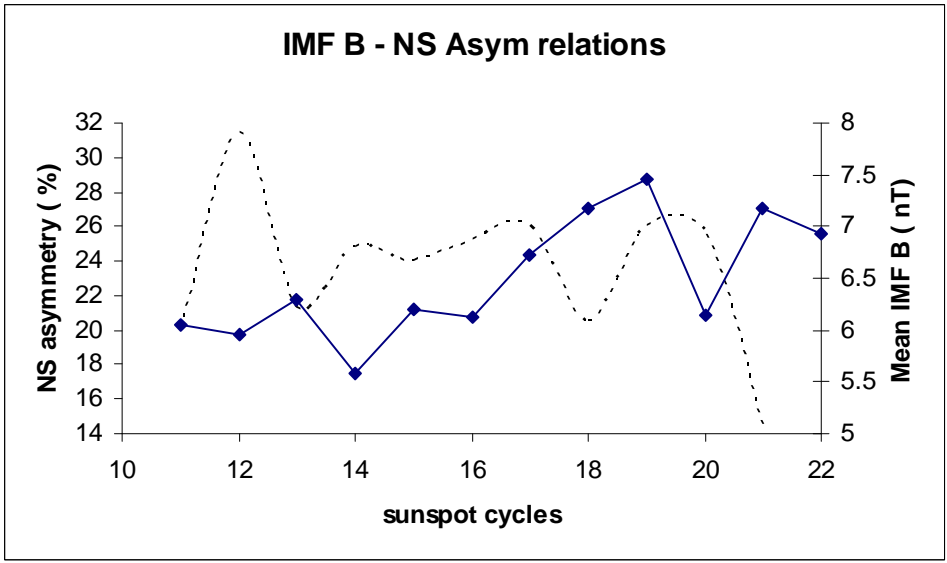

Fig 7

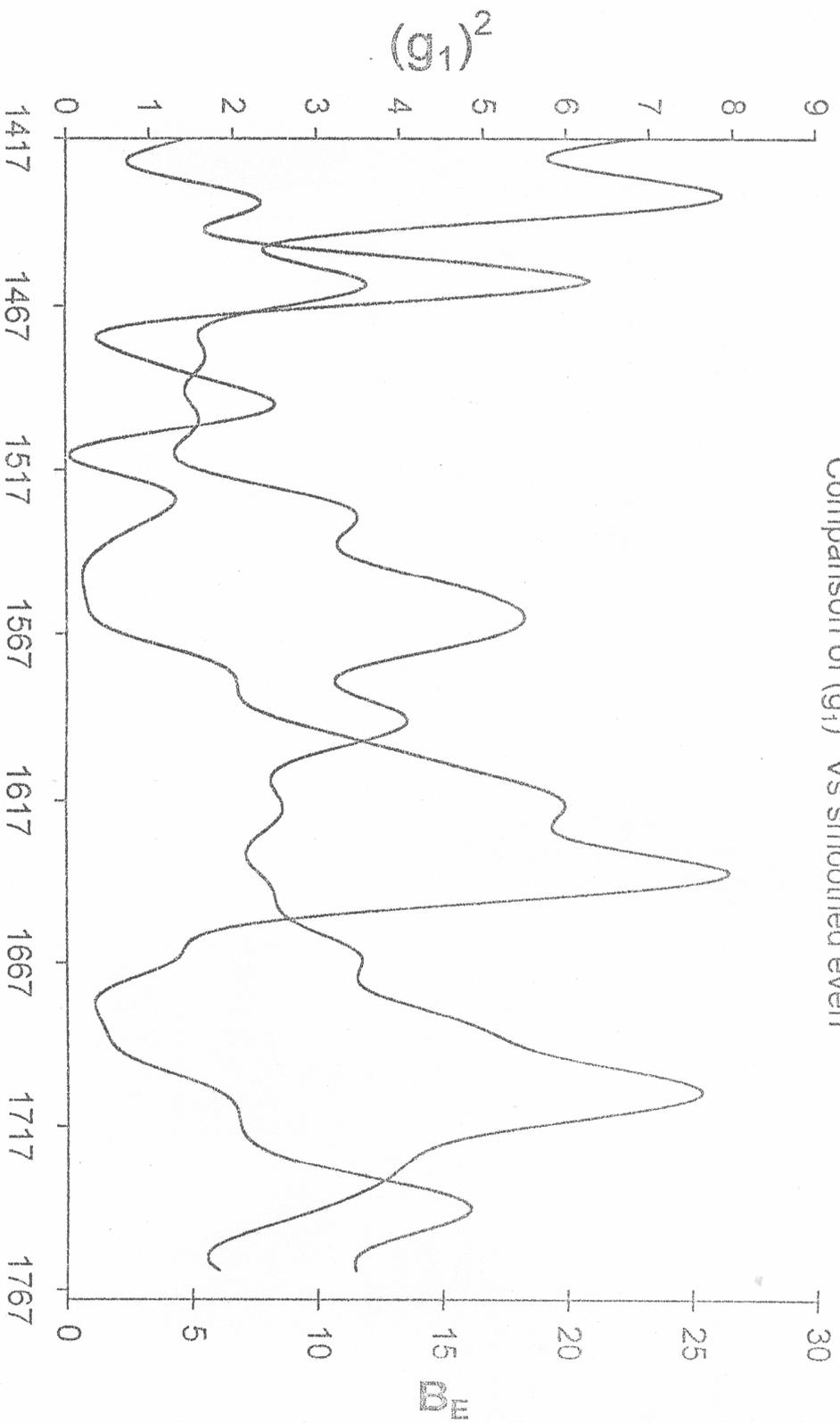